\documentclass[12pt,a4paper]{article}
\usepackage{graphicx}
\usepackage[T1]{fontenc}
\usepackage[utf8]{inputenc}
\usepackage{textcomp}
\usepackage[sc,osf]{mathpazo}
\usepackage{a4wide}  
\usepackage{latexsym,amsthm,amsfonts,amsmath,mathrsfs,amssymb}
\usepackage{dsfont}
\usepackage{accents}
\usepackage[nosort]{cite}
\usepackage{booktabs} 
\usepackage[unicode,implicit]{hyperref}
\hypersetup{%
  pdftitle    = {Black hole solutions in theories of supergravity}
  pdfkeywords = {history, supersymmetry supergravity, black holes, solutions, BPS},
  pdfauthor   = {Tomas Ortin},
  plainpages  = true,
  colorlinks  = true,
  citecolor   = blue,
  urlcolor    = red,
  linkcolor   = black
}
\newcommand{\hepth}[1]{{\tt
\href{http://www.arXiv.org/abs/hep-th/#1}{hep-th/#1}}}
\newcommand{\grqc}[1]{{\tt
\href{http://www.arXiv.org/abs/gr-qc/#1}{gr-qc/#1}}}

\newcommand{\arxiv}[1]{{\tt arXiv:\href{http://www.arXiv.org/abs/#1}{#1}}}
\newcommand{\doi}[1]{{\tt DOI:\href{http://dx.doi.org/#1}{#1}}}

\allowdisplaybreaks
\begin{document}

\begin{flushright}
\small
IFT-UAM/CSIC-24-138\\
\texttt{arXiv:2412.12020}\\
November 30\textsuperscript{th}, 2024\\
\normalsize
\end{flushright}

\vspace{1cm}

\renewcommand{\thefootnote}{\alph{footnote}}

\begin{center}

  {\Large {\bf Black hole solutions in theories of
      Supergravity}}\footnote{Invited contribution to the book \textit{Half a Century of
        Supergravity}, eds. A.~Ceresole and G.~Dall'Agata.}
 
\vspace{2cm}

{\sl\large Tom\'{a}s Ort\'{\i}n}\footnote{Email:
  {\tt tomas.ortin[at]csic.es}}

\vspace{2cm}

{\it Instituto de F\'{\i}sica Te\'orica UAM/CSIC\\
  C/ Nicol\'as Cabrera, 13--15,  C.U.~Cantoblanco, E-28049 Madrid, Spain}

\vspace{2cm}

{\bf Abstract}
\end{center}
\begin{quotation}
  {\small Supergravity has played a very important role in many advances and
    developments in black hole physics. Here I will review the history of some
    of them. }
\end{quotation}

\newpage
\pagestyle{plain}

\setcounter{footnote}{0}

\renewcommand{\thefootnote}{\arabic{footnote}}

It is fair to say that most of what we know about black holes has been learned
from the classical solutions that describe them. In 1967, when Wheeler gave a
name to the new concept, only a few black-hole solutions where known: the
static Schwarzschild--Droste \cite{kn:Schwarzschild} and Reissner--Nordstr\"om
\cite{kn:Reissner} solutions and their higher-dimensional and cosmological
generalizations found by Tangherlini \cite{Tangherlini:1963bw}, and the
stationary Kerr--Newman solution \cite{Kerr:1963ud}. The only multi-center
solutions known were those found by Majumdar and Papapetrou in
Refs.~\cite{Majumdar:1947eu} and they were not identified as multi-black-hole
solutions until 1972 \cite{Hartle:1972ya}. More examples were needed to test
and extend our understanding of these mysterious objects, finding general
patterns in their behaviors. Due to the uniqueness of black-hole solutions for
given values of the conserved charges, it was necessary to go beyond the
Einstein--Maxwell (EM) theory, considering more matter fields, more diverse
couplings and, eventually, higher-order terms. The possibilities were
virtually infinite, though, and a guiding principle was required in order to
select the theories of interest.

Since its inception, Supergravity offered such a guiding principle,
constraining the allowed matter and couplings, providing at the same time a
new and powerful tool that could also be used to study well-known theories
(EM, for instance): local supersymmetry. Supergravity theories, combined with
the Kaluza--Klein (KK) principle, provided a primitive candidate to Theory of
Everything \cite{Duff:1986hr} or low-energy effective field theory
approximations to Superstring Theories and, since black holes are probably the
``hydrogen atoms'' of Quantum Gravity, the interest in obtaining and studying
this kind of solutions in theories of Supergravity only grew with time
\cite{Ortin:2015hya}. All the \textit{stringy black-hole solutions} known and
studied so far are, in the end, nothing but Supergravity black-hole
solutions. Supergravity effectively provides the spacetime face of Superstring
black-hole physics. On the other hand, independently of supersymmetry and
Superstring Theory, Supergravity, which can always be seen as nothing but
General Relativity (GR) coupled to particular matter fields, has stimulated
research in higher-dimensional theories and classical solutions, leading to
transcendent results of universal validity such as the so-called
\textit{attractor mechanism} and its consequences \cite{Ferrara:1997tw} (see
Section~\ref{sec-attractor}).

Here I will review the history of the construction and study of black-hole
solutions in Supergravity, restricting myself to the asymptotically-flat ones
and to my personal recollections which could well be unfairly limited, for
which I apologize to all those concerned. More complete references can be
found in the reviews
\cite{DAuria:1998emc,Mohaupt:2000mj,Andrianopoli:2006ub,Ferrara:2008hwa,Bellucci:2009qv,Ortin:2015hya,Ortin:2016xvk}.

\section{The early history}

Supersymmetry is a spacetime symmetry. Thus, it is expected to introduce
modifications in our understanding of the gravitational interaction. These
modifications are due to the presence of matter (in the GR sense) fields in
the theory and should be specially relevant in black-hole spacetimes. Thus,
the search for supergravity black-hole solutions (or for the effect of
supersymmetry on known black-hole solutions) started immediately after the
construction of the first and simplest supergravity theory, whose matter
consisted only on one gravitino field. The effect of finite, local
supersymmetry transformations on the Schwarzschild and Reissner--Nordstr\"om
(RN) black-hole solutions was first studied in Ref.~\cite{Baaklini:1977bx},
but the solutions obtained there are gauge-equivalent to the original, purely
bosonic ones.  Furthermore, it was unclear how to analyze the geometry of a
metric with fermionic bilinears (\textit{i.e.} with supernumbers).  It was
conjectured in Ref.~\cite{Cordero:1978ud} that these were the only black-hole
solutions of the theory. Gueven proved this conjecture formulating a
``no-superhair theorem'' for spin-$3/2$ fields in Kerr black-hole background
in pure, ungauged, $\mathcal{N}=1,d=4$ supergravity \cite{Gueven:1980be}. This
result was soon extended to Kerr-Newman black holes in pure, ungauged,
$\mathcal{N}=2,d=4$ supergravity \cite{Aichelburg:1981rd}.

This no-superhair theorem could have sealed the fate of this line of work, but
it turned out to have an important exception: the extremal, (supersymmetric)
RN black hole may admit spin-$3/2$ hair \cite{Gueven:1982tk}. A most elegant
and interesting solution with non-trivial gravitini that reduces to the
extremal RN black hole when the gravitini vanish was constructed in
Ref.~\cite{Aichelburg:1983ux} and this work and its multicenter extensions
\cite{kn:Embacher1984} remains largely ignored due to the lack of a good
physical understanding of classical gravitini fields and, more generally, of
the supernumbers that enter the metric and Maxwell field components.

These difficulties led to the search for purely bosonic black-hole
solutions. Due to the uniqueness theorems, this search could only give really
new black holes for supergravities with $\mathcal{N}>2$ or $d>4$ or with
matter supermultiplets. This search was initiated by Gibbons in the context of
pure, ungauged $\mathcal{N}=4,d=4$ supergravity \cite{Gibbons:1982ih} and
widened to more general situations and extended objects in
Ref.~\cite{Gibbons:1987ps}.  These solutions played a very important role in
subsequent developments when they were rediscovered and reinterpreted as
solutions of the String Theory effective field theory action
(\textit{i.e.~stringy black holes}) \cite{Garfinkle:1990qj,Kallosh:1992ii}.

Simultaneously, the realization that all the solutions of the EM theory are
bosonic solutions of pure, ungauged $\mathcal{N}=2,d=4$ Supergravity suggested
the possibility of using supergravity techniques in the EM theory, such as the
Witten--Nester--Israel construction \cite{Witten:1981mf} used to prove the
positivity of mass, which is closely related to Supergravity
\cite{Hull:1983ap} as suggested by Deser and Grisaru in
Refs.~\cite{Deser:1977hu}. In $\mathcal{N}$-extended Supergravity, these
techniques lead to more restrictive bounds: the so-called
\textit{supersymmetry bounds},\footnote{Although they are very often referred
  to as ``Bogomol'nyi--Prasad--Sommerfield (BPS) bounds,''
  \cite{Bogomolny:1975de} not all BPS bounds are related to supersymmetry
  and, in this instance, supersymmetry bounds is a more precise name.} that
involve the $[\mathcal{N}/2]$ skew eigenvalues of the central charge

\begin{equation}
  M \geq |\mathcal{Z}_{1}|\,,
  \,\,\,\,\,
  \ldots
  \,\,\,\,\,,
  M \geq |\mathcal{Z}_{[\mathcal{N}/2]}|\,.
\end{equation}

\noindent
(Just one in $\mathcal{N}=2,d=4$ theories). In pure, ungauged,
$\mathcal{N}=2,d=4$ Supergravity, this bound coincides with the extremality
bound of RN black holes \cite{Gibbons:1982fy}, a fact later generalized to the
black holes of ungauged, $\mathcal{N}=2,d=4$ Supergravity coupled to vector
supermultiplets and to pure, ungauged, $\mathcal{N}=4,d=4$ Supergravity
\cite{Kallosh:1992ii}. The bound is saturated when some supersymmetries are
``restored'' or ``preserved.'' This is an important topic that deserves a more
detailed explanation.

Most field configurations (not necessarily solutions) of any theory are not
invariant under any of the symmetries of the theory. They ``break'' them
all. Some, though, are invariant under a, typically finite-dimensional, subset
of transformations which generate symmetries said to remain ``unbroken'',
``preserved'' or to be ``restored'' by the field configuration and which span a
Lie group of symmetries. The simplest example is the isometry group of a
metric, generated by its Killing vectors.

Bosonic field configurations of Supergravity theories may also preserve some
supersymmetries generated by a finite-dimensional space of spinorial
supersymmetry transformation parameters called, by analogy, \textit{Killing
  spinors}.  Field configurations admitting Killing spinors are called
\textit{supersymmetric}.\footnote{They are often called ``BPS'' but not all
  BPS configurations are supersymmetric. BPS configurations are characterized
  by the first-order differential equations they satisfy and the (BPS) bounds
  they saturate. Supersymmetric configurations have these two properties and
  are BPS.}  Together, the Killing vectors and spinors span a symmetry Lie
superalgebra \cite{Figueroa-OFarrill:1999klq}.

In purely bosonic field configurations ($F=0,B\neq 0$), the Killing spinors
$\epsilon$ solve the \textit{Killing spinor equations} (KSEs)

\begin{equation}
  \delta_{\epsilon}F
  =
  0\,,
\end{equation}

\noindent
for all the fermions $F$ of the theory. (Since the supersymmetry
transformations of the bosonic fields $B$ are proportional to the fermionic
ones $F$, $\delta_{\epsilon}B=0$ automatically.)

Apart from saturating supersymmetry bounds and satisfying the first-order
KSEs, supersymmetric configurations enjoy many other interesting properties
like the vanishing of their Hawking temperature \cite{Gibbons:1984kp} and of
some of the possible quantum corrections (depending on the theory and on the
amount of supersymmetry preserved) and satisfy the off-shell relations between
the equations of motion of the bosonic fields $\mathbf{E}_{B}$ known as
\textit{Killing spinor identities} (KSIs) \cite{Kallosh:1993wx}

\begin{equation}
\sum_{B}  \mathbf{E}_{B}\frac{\delta \delta_{\epsilon}B }{\delta F}
  =
  0\,,
\end{equation}

\noindent
which can be used instead of the integrability conditions of the KSEs
\cite{Bellorin:2005hy}. As a consequence, supersymmetric solutions and, in
particular, supersymmetric black-hole solutions, are easier to find because
one has to solve less independent equations of motion, just as staticity and
spherical symmetry lead to simpler ansatzs and less independent equations of
motion in GR. Furthermore, the supersymmetry bounds seem to (always?) imply
that several supersymmetric black holes can coexist in stationary equilibrium
(a phenomenon initially called ``antigravity'' \cite{Scherk:1979aj}), and one
can expect to find ``multi-center'' supersymmetric solutions which generalize
Majumdar and Papapetrou's (MP) \cite{Majumdar:1947eu}, although it is always
necessary to prove their regularity.

Thus, it is not surprising that a lot of effort has been invested over the
years in the construction and study of supersymmetric black-hole solutions, as
we are going to describe next.

\section{Supersymmetric black holes}
\label{eq:susybhs}

As a general rule, supersymmetric black holes are particular members of larger
families of Supergravity black-hole solutions\footnote{It should always be
  kept in mind that all the classical solutions of GR in vacuum are
  Supergravity solutions. Likewise, all the classical solutions of the EM
  theory are also solutions of pure, ungauged $\mathcal{N}=2,d=4$
  Supergravity. However, the same metric, with different identifications of
  the Maxwell field with Supergravity fields, can be \textit{embedded} as a
  Supergravity solution in many different theories and ways, sometimes, but
  not always, related by dualities. Different embeddings, specially if they
  are inequivalent under duality, may have different physical properties,
  preserving different amounts of supersymmetry and having different $\alpha'$
  corrections in String Theory.} with particular values of the physical
parameters (conserved charges and moduli) that characterize them. Those
particular values satisfy a supersymmetry bound and typically allow for an
event horizon. At the beginning they were identified by setting up the KSEs
for the whole family of solutions and showing that these KSEs admitted
solutions for those values of the parameters. In this way, the supersymmetric
members of the family of black-hole solutions found by Gibbons in
Ref.~\cite{Gibbons:1982ih} (some regular, some not) were identified in
Ref.~\cite{Kallosh:1992ii}. Also, an empirical relation between the amount of
supersymmetry preserved and the regularity of the supersymmetric black hole
solution was found: solutions preserving $1/4$ of the total
$\mathcal{N}=4,d=4$ supersymmetries had a regular horizon with finite area
(and, hence, finite Bekenstein--Hawking (BH) entropy) similar to that of the
extremal RN black hole while those preserving $1/2$ did not. Understanding
this fact requires an interpretation of those solutions as the duals of
backgrounds on which strings can be quantized leading to an appropriate
density of states (see Section~\ref{sec-stringbhs}).

The families known at the time of these developments were not the most general
Supergravity could admit. The most general supersymmetric black-hole solution
of a given Supergravity theory would be characterized by an electric and a
magnetic charge for each Abelian vector field and a \textit{modulus} (the
asymptotic value at spatial infinity) for each scalar field. The mass would be
determined by those parameters through the supersymmetry bound and the angular
momentum, in 4 dimensions, would always give rise to a naked singularity
\cite{Bellorin:2006xr} (but not in 5 \cite{Gauntlett:1998fz}). Some of the
independent parameters could be generated by performing global duality
transformations as in Ref.~\cite{Shapere:1991ta} and the most general solution
could be obtained starting with a \textit{generating solution}. The generating
solution for pure, ungauged, static, $\mathcal{N}=4,d=4$ black holes was
constructed in Ref.~\cite{Ortin:1992ur} and the most general supersymmetric
and non-supersymmetric solutions were explicitly constructed in
Refs.~\cite{Bergshoeff:1996gg} and \cite{Lozano-Tellechea:1999lwm},
respectively, using duality. Similar methods were employed including the
coupling to matter vector supermultiplets by Sen, Cveti\v{c} and Youm in
Refs.~ \cite{Sen:1994eb}.  For maximal Supergravities in $4<d<9$, a generating
solution was given in Ref.~\cite{Cvetic:1996zq} (see also
Ref.~\cite{Bertolini:1999je} for $\mathcal{N}=8,d=4$ and the proposal in
Ref.~\cite{Ortin:2012gg} for an explicit, completely general extremal
solution).

Although these methods were not the most efficient ones (see below), they led
to the important realization that physical quantities depending on the metric
alone (entropy, mass, temperature...) and physical laws such as the laws of
black-hole mechanics must be duality invariant. Both sides of the
supersymmetry bounds

\begin{equation}
  \label{eq:M=Z}
M= |\mathcal{Z}(\phi_{\infty},q,p)|\,, 
\end{equation}

\noindent
where $\phi_{\infty},q,p$ stand for all the moduli, electric and magnetic
charges a black hole can have, must be invariant, and the right-hand side can
eventually be determined by dimensional and group-theoretical arguments. The
BH entropy

\begin{equation}
S = S(q,p)\,,  
\end{equation}

\noindent
may also be determined by the same methods. Notice that we have not included
the moduli in the argument of the entropy function. Their absence in the
entropy of extremal, supersymmetric and non-supersymmetric black holes is a
major Supergravity-inspired result that can be proven using the
\textit{attractor mechanism} and will be discussed later (see
Section~\ref{sec-attractor}). From a more general perspective, these ideas and
developments allowed us to investigate the entropy of black holes directly
from the action, using its field content and (super)symmetries, without having
to construct the full solution.

The invariance of the black-hole's mass, temperature and entropy follows from
the invariance of the Einstein-frame metric. Supersymmetric solutions are
written in terms of a set of functions $H^{M}$ that satisfy simple equations
(Laplace equations in many cases). Since gauge and scalar fields are written
in terms of those functions, the $H^{M}$s must also transform under duality
but, then, it is clear that they must enter the metric in duality-invariant
combinations. Determining how these building blocks of the supersymmetric
solutions enter the different fields of the solutions is an important but
difficult task. As we will discuss in the next section, the structure of the
supersymmetric solutions in terms of the $H^{M}$s can also be used to
construct non-extremal (hence, non-supersymmetric solutions). We will give an
example of this essential property of Supergravity black-hole solutions in
Section~\ref{sec-attractor}.

If one is searching for supersymmetric black holes, the best strategy would be
to exploit the KSEs they satisfy. Tod pioneered this approach in
Ref.~\cite{Tod:1983pm} where he managed to characterize the most general
solutions of pure, ungauged, $\mathcal{N}=2,d=4$ Supergravity admitting at
least one\footnote{As a matter of fact, in this theory, all the supersymmetric
  solutions admit a (real) 4-dimensional space of Killing spinors
  (\textit{i.e.}~$1/2$ of the maximal possible number). The presence of matter
  or gaugings allows for solutions with less supersymmetry.}  Killing spinor
$\epsilon$ using the Newman--Penrose formalism. He found that they fall in two
(non-disjoint) classes depending on whether the vector bilinear
$\bar{\epsilon}\gamma^{a}\epsilon$ is timelike or null when $\epsilon$ is
treated as commuting instead of anti-commuting. In the timelike class one
finds the Perj\'es--Israel--Wilson (PIW) solutions \cite{Perjes:1971gv}, to
which the Majumdar--Papapetrou (and, hence, the extremal RN) family
belongs. In other theories, in the timelike class, one finds generalizations
of this family.\footnote{The generalizations of the PIW solutions that arise
  in more general theories are very interesting: the only regular black-hole
  solutions in the PIW class are the static MP solutions \cite{Hartle:1972ya}
  but, in more general theories there are multi-black-hole solutions with
  angular momentum (but static horizons, though) \cite{Denef:2000nb}. For
  Abelian vector fields, the relative positions of these black holes are
  determined by the condition of absence of Misner strings and vanishing total
  NUT charge, but for non-Abelian gauge fields these restrictions disappear
  \cite{Meessen:2017rwm}.} The null class contains pp-waves and, in other
theories, also string-like solutions that we will not discuss here.

The Newman--Penrose formalism is not very popular in our community and it has
not been fully developed in dimensions other than $d=4$. Thus, it was only
applied to one more theory (pure, ungauged, $\mathcal{N}=4,d=4$ Supergravity)
by Tod himself \cite{Tod:1995jf}. A slightly different method, fully
equivalent to Tod's in $d=4$, was proposed and used in
Ref.~\cite{Gauntlett:2002nw} to characterize all the supersymmetric solutions
(and find many interesting ones)\footnote{Apart from the maximally
  supersymmetric solutions (Minkowski, the Robinson--Bertotti- (or
  Freund--Rubin)-type AdS$_{2}\times$S$^{3}$ and AdS$_{3}\times$S$^{2}$
  solutions that arise in the near-horizon limits of black-hole and string
  solutions, a 5-dimensional generalization of the 4-dimensional
  Kowalski--Glikman pp-wave previously found in Ref.~\cite{Meessen:2001vx}
  which arises as the Penrose limit \cite{kn:Penrose1976} of the former and a
  and a G\"odel-type solution) an embedding of the extremal RN--Tangherlini
  and a (regular!) rotating version of it which is a particular case of of the
  BMPV black hole \cite{Breckenridge:1996is}. The supersymmetric black ring of
  Ref.~\cite{Elvang:2004rt} was also shown to be a particular solution of this
  theory \cite{Gauntlett:2004qy}. The G\"odel solution truly works as a vacuum
  that admits asymptotically-G\"odel, charged, rotating, black-hole and
  black-ring solutions \cite{Ortin:2004af}. More recently, supersymmetric,
  asymptotically-flat black holes whose horizon has the topology of a lense
  space have been constructed and embedded in String Theory in
  Refs.~\cite{Kunduri:2014kja}. A classification of the possible near-horizon
  topologies of extremal black holes in different dimensions can be found in
  Ref.~\cite{Kunduri:2013gce}.} of minimal $\mathcal{N}=1,d=5$
Supergravity. This method is based on the construction of $p$-form bilinears
of the Killing spinor whose existence is assumed at the beginning. These
bilinears satisfy an algebra which is, ultimately, a rewriting of the relevant
Clifford algebra but provides a much more useful and insightful understanding
of the geometry of the supersymmetric solutions. The KSEs are translated into
first-order differential equations for the bilinear $p$-forms that determine
necessary conditions for a field configuration (not necessarily a solution) to
be supersymmetric. These conditions are subsequently checked to be sufficient
and, finally, one imposes the equations of motion that remain independent
after use of the KSIs \cite{Kallosh:1993wx,Bellorin:2005hy}, which can also be
rewritten in terms of the bilinears.

This method was soon applied to encompass other matter couplings of
$\mathcal{N}=1,d=5$ Supergravity: to gauged minimal Supergravity in
Ref.~\cite{Gauntlett:2003fk}, to the coupling to Abelian vector
supermultiplets in \cite{Gauntlett:2004qy} and to the gauging of those Abelian
vector multiplets \cite{Gutowski:2004yv}. Hypermultiplets, tensor multiplets,
and general, non-Abelian gaugings were considered in
Refs.~\cite{Bellorin:2006yr}. In absence of gaugings, hypermultiplets cannot
be active in asymptotically-flat black holes because they would be primary
hair \cite{Coleman:1991ku}. Abelian gaugings lead to asymptotically-AdS black
holes but non-Abelian gaugings have been used to construct, apart from global
monopole solutions, asymptotically-flat, regular black holes which are not
just the embedding of an Abelian solution \cite{Meessen:2015enl}.

The bilinear method was also applied to $\mathcal{N}=2,d=4$ Supergravities:
gauged, minimal \cite{Caldarelli:2003pb}, ungauged with vector supermultiplets
\cite{Meessen:2006tu} (recovering and generalizing the results of
Refs.~\cite{Behrndt:1997ny,Denef:2000nb}) and hypermultiplets
\cite{Huebscher:2006mr}, and gauged with Abelian \cite{Cacciatori:2008ek} and
non-Abelian \cite{Huebscher:2007hj,Meessen:2012sr} gaugings. These results
provided algorithms to construct asymptotically-AdS and non-Abelian black
holes, respectively. The relation between 5- and 4-dimensional non-Abelian
black holes of Supergravity theories with 8 supercharges is akin to the
relation between 4-dimensional self-dual Yang--Mills instantons and
3-dimensional monopoles satisfying the Bogomol'nyi equation, both in their
gauge fields (obviously) and in their spacetime
geometry.\footnote{5-dimensional timelike supersymmetric solutions include a
  hyperK\"ahler ``base'' space which must be of Gibbons--Hawking type to allow
  for its dimensional reduction.} The monopole fields that can be used to
construct regular supersymmetric black holes is very large, though, and one
can also construct \textit{coloured black holes} \cite{Meessen:2008kb}. It is
truly remarkable that supersymmetry made possible the construction of
analytical solutions with non-Abelian fields since, until then, all the
solutions found were either embeddings of Abelian black holes or numerical
solutions.

Still in 4 dimensions, Tod's work in $\mathcal{N}=4,d=4$ \cite{Tod:1995jf} was
revisited using the bilinear approach in Ref.~\cite{Bellorin:2005zc} and, in
Ref.~\cite{Meessen:2010fh} the general form of the black hole of all extended,
ungauged, Supergravities coupled to vector supermultiplets was given. While
this result may be seen as the final solution to the problem of finding the
most general 4-dimensional, supersymmetric, asymptotically-flat, black holes
with Abelian charges the solutions are given in terms of objects that obey
constraints which are, actually, very difficult to solve, although a solution
was found for the $\mathcal{N}=8$ case in Ref.~\cite{Ortin:2012gg}. Since many
of the 4-dimensional extended Supergravities have consistent $\mathcal{N}=2$
truncations, one may hope to solve them using the results obtained for those
theories \cite{Denef:2000nb,Meessen:2006tu}. However, here the solutions
depend on objects that satisfy complicated equations: the so-called
\textit{stabilization equations} that relate the real and imaginary parts of a
symplectic vector of complex functions. It is worth remembering the remarkable
solution found by Shmakova for cubic models that originate in Calabi--Yau
compactifications of $\mathcal{N}=2,d=10$ theories \cite{Shmakova:1996nz}.

Finally, similar ``classification of all the supersymmetric solutions'' works
have been carried out in minimal \cite{Gutowski:2003rg} and non-minimal
\cite{Cariglia:2004kk} $\mathcal{N}=(1,0),d=6$ Supergravity.

\subsection{Non-supersymmetric black holes in Supergravity}
\label{sec-nonsusy}

Not all extremal black holes are supersymmetric. Actually, it can be argued
that most of them are not: in many cases (but not always \cite{Khuri:1995xq}),
supersymmetric black-hole solutions can be transformed into non-supersymmetric
by just flipping the sign of one of its charges \cite{Ortin:1996bz}. Since
dualities preserve the number of unbroken supersymmetries, non-supersymmetric
black holes belong to duality orbits disjoint from the supersymmetric ones and
saturate different, non-supersymmetric, BPS bounds in which the relevant skew
eigenvalue of the central charge matrix Eq.~(\ref{eq:M=Z}) is replaced by a
\textit{fake central charge} \cite{Ceresole:2007wx}

\begin{equation}
  M =|\mathcal{Z}_{\rm fake}(\phi_{\infty},p,q)|\,,  
\end{equation}

\noindent
and the entropy function $S(p,q)$ is a different duality invariant, still
independent of the moduli because the attractor mechanism works for all
extremal black holes.

Non-supersymmetric extremal black holes share many of the properties of their
supersymmetric cousins. In particular, their BPS bounds can be related to
no-force conditions and typically there are non-supersymmetric
multi-black-hole solutions.

Non-extremal black holes (always non-supersymmetric) are much more interesting
objects since they are more similar to the real-world beasts we would like to
understand (ideally, the Kerr black hole). Perhaps not surprisingly they are
more difficult to understand. Obtaining the classical solutions that represent
them is also more difficult and they have not been explored as systematically
as the extremal ones. Some tricks like deforming the metric of the extremal
black hole that has the same charges with a \textit{blackening factor} seem to
work in many cases, but there is a more interesting proposal based on the
observation made above on the decomposition of the supersymmetric solutions in
terms of functions $H^{M}$ that satisfy simple equations and which transform
under duality in a well-defined way (\textit{e.g.}~as symplectic vectors in
$\mathcal{N}=2,d=4$ supergravities). This property seems to hold true for
non-extremal solutions as well \cite{Mohaupt:2009iq} as the black-hole
solutions of pure, ungauged, $\mathcal{N}=4,d=4$ Supergravity
\cite{Lozano-Tellechea:1999lwm} and of some models of ungauged
$\mathcal{N}=2,d=4$ and $\mathcal{N}=1,d=5$ Supergravity coupled to vector
multiplets \cite{Galli:2011fq} show.  This, is actually, the basis of the
so-called \textit{H-FGK formalism} \cite{Meessen:2011aa}.  Obviously, there
are differences: the equations satisfied by the building blocks are different,
more complicated and they enter the metric in a slightly different way, but
the structure of the solutions, imposed by the duality of the theory, is the
same.

\section{The attractor mechanism}
\label{sec-attractor}

The attractor mechanism is one the most fascinating properties of extremal
black holes discovered in the context of Supergravity, more precisely, in the
context of ungauged $\mathcal{N}=2,d=4$ Supergravity coupled to vector
multiplets \cite{Ferrara:1995ih} and for static, supersymmetric, black
holes. It was soon realized that it is a property of all the 4-dimensional
extremal black-hole \cite{Ferrara:1997tw} (and also higher-dimensional
black-hole and $p$-black-brane \cite{deAntonioMartin:2012bi}) solutions of any
Supergravity-like theory containing scalars coupled to vector (or
$(p+1)$-form) potentials. In this section I am going to review the basic facts
and the most important consequences\footnote{More details can be found in the
  reviews
  \cite{Ortin:2015hya,DAuria:1998emc,Andrianopoli:2006ub,Ferrara:2008hwa,Bellucci:2009qv,Ortin:2016xvk,Pioline:2006ni}
  and references therein.}  but I would like to start by stressing that the
description of the mechanism as the simple fact that the value of the scalars
on the horizon of any static, extremal, black hole is always independent of
its value at infinity is an oversimplification which, furthermore, is
generically not true. I will only consider the 4-dimensional, black-hole case
studied in Ref.~\cite{Ferrara:1995ih}, which applies to Supergravity-like
theories with real scalar fields $\phi^{i}$ and vector fields
$A^{\Lambda}{}_{\mu}$ with field strengths
$F^{\Lambda}{}_{\mu\nu}=2\partial_{[\mu}A^{\Lambda}{}_{\nu]}$ coupled to
gravity as prescribed by the action

\begin{equation}
\label{eq:genericd4action2}
S
= 
\int
 d^{4}x
\sqrt{|g|}
\left\{
R
+
\mathcal{G}_{ij}\partial_{\mu}\phi^{i}\partial^{\mu}\phi^{j}
+
2\Im\mathfrak{m}\mathcal{N}_{\Lambda\Sigma}
F^{\Lambda\, \mu\nu}F^{\Sigma}{}_{\mu\nu}
-
2\Re\mathfrak{e}\mathcal{N}_{\Lambda\Sigma}
F^{\Lambda\, \mu\nu}\, \star F^{\Sigma}{}_{\mu\nu}
\right\}\,.
\end{equation}

For static, spherically-symmetric black-hole solutions, one can use the
following ansatz for the metric  \cite{Breitenlohner:1987dg}:

\begin{equation}
\label{eq:generalbhmetric2}
ds^{2} 
 = 
e^{2U} dt^{2} - e^{-2U}
\left[\frac{r_{0}^{4}}{\sinh^{4}{(r_{0}\rho)}}d\rho^{2} 
+
\frac{r_{0} ^{2}}{\sinh^{2}{(r_{0}\rho)}}d\Omega^{2}_{(2)}\right]\,,
\end{equation}

\noindent
which depends on a single unknown function $U(\rho)$ and on the
\textit{non-extremality parameter} $r_{0}$ related to the temperature $T$ and
entropy $S$ of the black holes described by it by

\begin{equation}
r_{0} = 2ST\,, 
\end{equation}

\noindent
so that it vanishes or extremal black holes.\footnote{This ansatz for the
  metric is the magic ingredient in the following results. The generalization
  needed to derive similar results in higher dimensions and for
  higher-$p$-branes was introduced in Ref.~\cite{deAntonioMartin:2012bi}.} The
Maxwell equations can be integrated (the magnetic and electric charges
$(Q^{M})= \left(
  \begin{array}{c}
    p^{\Lambda} \\ q_{\Lambda} \\
  \end{array}
\right)$ arise as integration constants) and, for scalars
$\phi^{i}=\phi^{i}(\rho)$ and for the function $U(\rho)$ the rest of the
original equations of motion reduce to those derived from the effective
(\textit{FGK}) mechanical action

\begin{equation}
\label{eq:effectiveaction}
S_{\rm eff}[U,\phi^{i}] 
= 
\int d\rho\, 
\left[
\dot{U}^{2}  
+\tfrac{1}{2}\mathcal{G}_{ij}\dot{\phi}^{i}  \dot{\phi}^{j}  
-e^{2U} V_{\rm bh}(\phi)
\right]\,,
\hspace{1cm}
\end{equation}

\noindent
where the so-called \textit{black-hole potential} $V_{\rm bh}(\phi)$ is
defined by

\begin{equation}
\label{eq:MofN}
  V_{\rm bh}(\phi)
  =
  -\tfrac{1}{2}\mathcal{Q}^{M} \mathcal{M}_{MN} \mathcal{Q}^{N}\,,
  \hspace{1cm}
\left(\mathcal{M}_{MN}\right)
\equiv 
\left(
  \begin{array}{cc}
I_{\Lambda\Sigma} +R_{\Lambda\Gamma}I^{\Gamma\Omega}R_{\Omega\Sigma}\,\,\,\,
& 
-R_{\Lambda\Gamma}I^{\Gamma\Sigma} 
\\
& \\
-I^{\Lambda\Omega}R_{\Omega\Sigma} 
& 
I^{\Lambda\Sigma}
\\
\end{array}
\right)\,.
\end{equation}

\noindent
The Hamiltonian is conserved and the original equations of motion imply that
it must take the value $r_{0}^{2}$ (\textit{Hamiltonian
  constraint})\footnote{It is worth stressing that we are actually dealing
  with a family of different actions for each set of charges. It is worth
  mentioning that different sets of charges may lead to equivalent FGK
  actions. Thus, the \textit{Freudenthal dual} of $Q^{M}$ leads to the same
  black-hole potential, with the same critical points and to black holes with
  and entropies \cite{Borsten:2009zy,Ferrara:2011gv,Galli:2012ji}.}

\begin{equation}
\dot{U}^{2}  
+\tfrac{1}{2}\mathcal{G}_{ij}\dot{\phi}^{i}  \dot{\phi}^{j}  
-e^{2U} V_{\rm bh}(\phi)
=
r_{0}^{2}\,.
\end{equation}

Under very mild assumptions it is possible to proof that, if
$\phi^{i}_{\rm h}$ are the constant values of the scalars over the event
horizon, they extremize the black-hole potential for given charges

\begin{equation}
  \label{eq:criticalpointsVbh}
  \left.\partial_{i}V_{\rm bh}\right|_{\phi= \phi_{\rm h}}
  =
  0\,,
\end{equation}

\noindent
and the of value of $V_{\rm bh}$ at this extremum essentially gives the
entropy

\begin{equation}
\label{eq:entropyandbhpotential}
S/\pi = -V_{\rm bh}(\phi_{\rm h},\mathcal{Q})\,.
\end{equation}

Furthermore, for each solution of Eqs.~(\ref{eq:criticalpointsVbh}) (each
\textit{attractor}\footnote{The name is associated to the interpretation of
  $\rho$ as a time parameter describing the evolution of the variables
  $U(\rho),\phi^{i}(\rho)$ from spatial infinity to the horizon.}
$\phi^{i}_{\rm h}$) there exists at least one extremal black-hole solution,
the so-called \textit{double extremal black hole}, with constant scalars (in
poarticular, the moduli $\phi^{i}_{\infty}=\phi^{i}_{\rm h}$) equal to the
attractor and mass given by

\begin{equation}
\label{eq:massdoubleextremal}
M^{2}_{\rm double\,\, extremal}
=
-V_{\rm bh}(\phi_{\rm h},\mathcal{Q})
=
S/\pi\,.
\end{equation}

\noindent
Other extremal black holes with the same attractor have different values of
the scalars at infinity and, therefore, non-constant scalars. Their masses are
given by

\begin{equation}
\label{eq:genericbound}
M^{2}
+  
\tfrac{1}{2}\mathcal{G}_{ij}(\phi_{\infty})\Sigma^{i}  \Sigma^{j}  
+V_{\rm bh}(\phi_{\infty},\mathcal{Q})
= 
r_{0}^{2}\geq 0\,,
\end{equation}

\noindent
where the $\Sigma^{i}$, the scalar charges, must be functions of the conserved
charges and moduli $\Sigma^{i} =\Sigma^{i}(\phi_{\infty},\mathcal{Q},M)$
(\textit{i.e.}~there is no primary scalar hair \cite{Coleman:1991ku}) if the
event horizon is regular \cite{Ballesteros:2023iqb}.\footnote{These scalar
  charges also occur in the first law of black hole mechanics, as shown in
  Ref.~\cite{Gibbons:1996af}.}

Eqs.~(\ref{eq:criticalpointsVbh}) relate the critical values of the black-hole
potential $\phi^{i}_{\rm h}$ to the charges. The solution is unique
$\phi^{i}_{\rm h}= \phi^{i}_{\rm h}(\mathcal{Q})$, independent of the moduli
$\phi^{i}_{\infty}$ (hence the name attractor) only when $V_{\rm bh}$ has no
flat directions, a property that depends on the values of the charges and
always holds for the supersymmetric case. When this happens,
Eq.~(\ref{eq:entropyandbhpotential}) guarantees that the entropy only depends
on the charges, which are quantized, and it is, therefore, a function of
integer numbers, a result that strongly calls for a microscopic interpretation
in terms of state counting.

For general values of the charges, though, $V_{\rm bh}$ has flat directions
and the attractors also depend on other continuous parameters that may include
the moduli. One may say that the attractor mechanism does not work in general,
but the crucial point is that even in these cases the entropy only depends on
the charges because of the very definition of flat direction
\cite{Sen:2005iz}.\footnote{Notice that Eq.~(\ref{eq:massdoubleextremal})
  indicates that this property also applies to the masses of double-extremal
  black holes.}

It is difficult to overstate the importance of this Supergravity-inspired but
rather general result that could well have been derived by classical
relativists long before (but was not).

When the action Eq.~(\ref{eq:genericd4action2}) is that of the bosonic sector
of a Supergravity theory, the black-hole potential is determined by functions
which play important roles in the theory. For instance, in $\mathcal{N}=2,d=4$
supergravity coupled to vector supermultiplets, which contain a complex scalar
$Z^{i}$, a vector field $A^{i}$ and a gaugino. The scalars parametrize a
special K\"ahler manifold with metric $\mathcal{G}_{ij^{*}}$ and the matter
vector fields together with the graviphoton are denoted by $A^{\Lambda}$. The
black-hole potential is a function of the absolute value of the central charge
$\mathcal{Z}$

\begin{equation}
\label{eq:centralchargeN2d4}
\mathcal{Z}(\mathcal{Q},Z,Z^{*}) 
\equiv 
\mathcal{V}_{M}\mathcal{Q}^{M}\,,  
\end{equation}

\noindent
where $\mathcal{V}_{M}$ is canonical covariantly-holomorphic symplectic
section that defines the theory\footnote{See, for instance
  Refs.~\cite{Andrianopoli:1996cm,Ortin:2015hya}.} by

\begin{equation}
\label{eq:N2d4bhpotential2}
-V_{\rm bh}(Z,Z^{*},\mathcal{Q}) 
= 
|\mathcal{Z}|^{2} 
+4\mathcal{G}^{ij^{*}}\partial_{i}|\mathcal{Z}|\partial_{j^{*}}|\mathcal{Z}|\,,
\end{equation}

\noindent
For supersymmetric extremal black holes
($M=|\mathcal{Z}(\phi_{\infty},\mathcal{Q})|\equiv |\mathcal{Z}_{\infty}|$)
\cite{Ferrara:1997tw}

\begin{equation}
S/\pi = |\mathcal{Z}_{\rm h}|^{2}
\hspace{1cm}
\left.\partial_{i}|\mathcal{Z}| \right|_{\rm h}= 0\,,
\end{equation}

\noindent
so that the supersymmetric attractors are critical points of the central
charge, which determines the entropy and the mass. Furthermore, 

\begin{equation}
\left.\partial_{i}\partial_{j^{*}} V_{\rm bh}\right|_{Z_{\rm h\, susy}} 
= 
\left. 2\mathcal{G}_{ij^{*}} V_{\rm bh} \right|_{Z_{\rm h\, susy}},
\hspace{1cm}
\left.\partial_{i}\partial_{j^{*}} |\mathcal{Z}|\right|_{Z_{\rm h\, susy}} 
= 
\left. \tfrac{1}{2}\mathcal{G}_{ij^{*}} |\mathcal{Z}| \right|_{Z_{\rm h\, susy}},
\end{equation}

\noindent
so that $V_{\rm bh}$ has no flat directions and there is a maximum of
$V_{\rm bh}$ there (which implies stability) and a minimum of $|\mathcal{Z}|$.

These properties do not hold in general for non-supersymmetric attractors
\cite{Goldstein:2005hq}.

Using Eq.~(\ref{eq:N2d4bhpotential2}) the FGK action
Eq.~(\ref{eq:effectiveaction}) can be rewritten, up to a total derivative, as
a sum of non-negative terms, \textit{\`a la} Bogomol'nyi
\cite{Bogomolny:1975de,Ferrara:1997tw}:

\begin{equation}
\label{eq:FGKactionBPSform}
S[U,Z^{i}] 
=
\int d\rho 
\left[ 
\left(\dot{U}\pm e^{U}|\mathcal{Z}|\right)^{2}
+\mathcal{G}_{ij^{*}}
\left(\dot{Z}^{i}\pm 2e^{U} \partial^{i}|\mathcal{Z}|\right)
\left(\dot{Z}^{*\, j^{*}}\pm 2e^{U} \partial^{j^{*}}|\mathcal{Z}|\right)  
\right]\,.
\end{equation}

The extrema of the FGK action (classical solutions) are configurations that
make each of these terms vanish, which happens when the following \textit{flow
  equations} are satisfied\footnote{We have selected the sign that gives a
  positive mass.}

\begin{equation}
\label{eq:BPSflowequations}
\dot{U}
 =  
 e^{U}|\mathcal{Z}|,
\hspace{1cm}
\dot{Z}^{i}
= 
 2e^{U} \partial^{i}|\mathcal{Z}|\,.
\end{equation}

These flow (or BPS) equations also imply the Hamiltonian constraint, but with
$r_{0}=0$ (extremal case) and can also be obtained by requiring unbroken
supersymmetry. In the near-horizon limit, these equations give the attractor
mechanism for the supersymmetric case.

There is no unique way of rewrite an action \textit{\`a la} Bogomol'nyi and,
in this case, it was shown in Ref.~\cite{Ceresole:2007wx,LopesCardoso:2007qid}
that the FGK action can be put in the form Eq.~(\ref{eq:FGKactionBPSform})
with $\mathcal{Z}$ replaced by \textit{fake central charges} associated with
non-supersymmetric attractors. In all these cases the ``evolution'' of the
scalars from spatial infinity to the horizon is driven by a single (fake)
central charge that appears in the flow equations, but there are more exotic
cases in which the behavior near the horizon is driven by one charge and at
spatial infinity by another charge. The H-FGK formalism
\cite{Mohaupt:2009iq,Meessen:2011aa}is specially suited to study these
phenomena. On the other hand, there are rewritings of the action \textit{\`a
  la} Bogomol'nyi for non-extremal black holes as well
\cite{Miller:2006ay}. For the relation between the first-order and
Hamilton-Jacobi formalisms in this context see,
\textit{e.g.}~Refs.~\cite{Skenderis:2006rr}.

\section{Stringy black holes}
\label{sec-stringbhs}

Research on stringy black holes started at a moment in which the worldsheet
and CFT approaches to Superstring Theory had come to a standstill and were
part of a new effort to explore it from the spacetime point of
view.\footnote{This standstill arrived more or less at the time of the defense
  of my PhD thesis on string multiloop amplitudes and string
  thermodynamics. Many people trained and working on those approaches left the
  field. As usual, big revolutions (string revolutions or other) are preceded
  by big crises which are much less talked about.  In that sense, the timing
  of my arrival in Stanford thanks to a postdoctoral Spanish government grant
  was perfect, and very fortunate.
  Refs.~\cite{Garfinkle:1990qj,Shapere:1991ta} and a few others had just been
  published.  There, I met Prof.~Kallosh, who was also looking for new lines
  of research and who introduced me to this field and to her collaborators,
  Profs.~Bergshoeff and Van Proeyen, from which I learned much of what I know
  on the spacetime (Supergravity) approach.} As we have stressed in the
introduction, Supergravity provided spacetime Lagrangians that describe the
low-energy dynamics of the massless modes of the theory even when those
effective theories had not been derived from string amplitudes or from the
quantum-mechanical conformal invariance of the worldsheet action because local
supersymmetry is a very powerful constraint (see,
\textit{e.g.}~Ref.~\cite{Bergshoeff:1989de}). Moreover, those Lagrangians had
very mysterious global symmetries \cite{Gaillard:1981rj} which could be used
as solution-generating transformations.

The key to exploit Supergravity as a tool to explore String Theory is the
dictionary that relates the spacetime Supergravity fields (metric, dilaton,
Kalb--Ramond, Ramond--Ramond and Yang--Mills fields in 10 dimensions) to the
massless modes in the string spectrum and the classical solutions of the
Supergravity theories to intersecting \cite{Papadopoulos:1996uq} massive
extended solitonic objects wrapped around compact dimensions
\cite{Duff:1994an}. Some important entries in this dictionary were the
interpretation of the scalar field of $\mathcal{N}=2A,d=10$ Supergravity with
the type~IIA Superstring dilaton and with the KK scalar coming from the
dimensional reduction of 11-dimensional supergravity on a circle
\cite{Witten:1995ex}, the identification of the Ramond--Ramond fields, sourced
by D-branes, in $\mathcal{N}=2,d=10$ Supergravities \cite{Polchinski:1995mt}
and the relation between the SL$(2,\mathbb{R})$ global symmetry of
$\mathcal{N}=2B,d=10$ Supergravity compactified on a circle with the modular
group of 11-dimensional Supergravity compactified on a torus
\cite{Bergshoeff:1995as}.  With this dictionary one can translate the
mysterious global symmetries discovered in Supergravity into, often new and
unexpected, String Theory \textit{dualities} \cite{Hull:1994ys}. Many of these
dualities involve non-perturbative states which are associated with classical
Supergravity solutions and they would have been impossible to discover using
the standard worldsheet methods.

The (Supergravity) effective field theory description of Superstring Theory
has a range of validity and clear limitations: the curvatures involved must be
smaller than $\alpha'$ and one must stay away from limits in which
symmetry-enhancing phenomena associated to new massless modes take
place. Still, even with all those caveats, it has been an extremely useful
description, specially of black holes.

Some of the earliest research on stringy black holes was based on inconsistent
truncations of the effective actions\footnote{To this day, the so-called
  dilaton--Gauss-Bonnet model is often described as a model whose black-hole
  solutions can be understood as stringy, but this model is an inconsistent
  truncation of the string effective action \cite{Cano:2021rey}.} and their
relevance is, at the very least, debatable. The first black-hole solutions
which were labeled ``stringy'' were the Garfinkle--Horowitz--Strominger
solutions of a very simple 1-parameter family of models nowadays known as
Einstein--Maxwell--Dilaton (EMD) model. This model describes a scalar field
non-minimally coupled to the Maxwell field with an strength controlled by the
parameter, $a$ \cite{Garfinkle:1990qj} and it (and its solutions) were
particular cases of a similar, Supergravity-inspired model constructed by
Gibbons and Maeda \cite{Gibbons:1987ps} in which the Maxwell field was
replaced by a $(p+1)$-form. In spite of the claims, only for very few values
of the parameter ($a=0,1/\sqrt{3},1,\sqrt{3}$) the model actually was a
consistent truncation of a Supergravity theory and of an effective string
theory action and the solutions were stringy black holes. The impact of their
discovery at the time cannot be overstated and it brought attention to another
family of black-hole solutions found much earlier by Gibbons in pioneering
work in the context if pure, ungauged, $\mathcal{N}=4,d=4$ Supergravity
\cite{Gibbons:1982ih}, a theory that can be obtained from the 10-dimensional
effective action of the heterotic superstring by compactification in a T$^{6}$
followed by the consistent truncation of the matter vector supermultiplets.

Thus, the stringy nature of those black holes was clear, and knowing their
embedding in a Supergravity theory made it possible for their possible
unbroken supersymmetries to be searched for. The (by now well-known) relation
between extremality and existence of unbroken supersymmetries was confirmed
and a mysterious relation between the amount of unbroken supersymmetry and the
regularity of the event horizon was also uncovered in
Ref.~\cite{Kallosh:1992ii}. This relation is now understood as a consequence
of the presence of more extended objects breaking supersymmetry in the
associated String Theory background. As a general rule, several of these
objects (breaking more supersymmetry) are needed for the density of quantum
string states living in that background to grow with the energy as needed to
give rise to a Von Neumann entropy that can be related to the
Bekenstein--Hawking (BH) entropy of the corresponding black holes.

A big step towards this discovery was the study of the 4-dimensional ``STU
model'' \cite{Duff:1995sm}, which can be obtained by compactification and
consistent truncation from several string effective field theories
(supergravities). It contains several vector fields that can be associated to
different extended objects and several scalars related by duality
symmetries. The solutions of the EMD model with the values of the parameter
$a$ mentioned above can be embedded in the STU model (hence, in String Theory)
for different identifications of charges. In particular, the stringy 4-charge,
4-dimensional Maldacena--Strominger--Johnson--Khuri--Myers
\cite{Maldacena:1996gb} black hole is a solution of the STU model that reduces
to known black-hole solutions of the EMD model (including the
Reissner--Nordstr\"om one) for different identifications of the charges.

As it is well known, the main revolution in this field was the famous
Strominger and Vafa paper Ref.~\cite{Strominger:1996sh} relating the
classical, macroscopic, BH entropy of a 5-dimensional stringy, static,
charged, extremal, asymptotically-flat supersymmetric black hole
(\textit{i.e.}~one quarter of the area of the horizon) to the density of
states of a string theory defined on a background containing certain D-branes,
whose Ramond--Ramond charges are essentially the black-hole's charges. This
result was quickly extended to 4-dimensional black holes
\cite{Maldacena:1996gb}, to 5-dimensional rotating black holes
\cite{Breckenridge:1996is}, and to generic 5-dimensional black holes obtained
by compactifications of 11-dimensional Supergravity on a Calabi--Yau 3-fold
\cite{Maldacena:1997de}, all of them extremal and supersymmetric and all of
them valid at zeroth order in $\alpha'$.\footnote{These momentous results were
  also shocking to some of us because they contradicted the results of
  Refs.~\cite{Gibbons:1994ff} obtained using the Euclidean path integral
  approach according to which the entropy of extremal black holes was
  identically zero when calculated directly at extremality and not as a limit,
  as in Ref.~\cite{Kallosh:1992wa}. This discrepancy between the String Theory
  and Euclidean approaches to Quantum Gravity nearly killed the latter which
  was dimmed invalid, at the very least.\footnote{Maldacena 1996, private
    communication.}. Now that research on String Theory seems to be reaching a
  new standstill, the Euclidean approach has been resurrected and used to show
  that, taking into account 1-loop corrections to the saddle-point
  approximation, one can show that the entropy of extremal non-supersymmetric
  black holes is indeed zero but that of the supersymmetric ones is not
  \cite{Iliesiu:2020qvm}. This is an ongoing and very interesting research
  topic.}

From this point onwards, the story of the zeroth-order\footnote{This is often
  called ``the Supergravity approximation'', although, when one has control
  over the local supersymmetry of the higher-order corrections, as in
  Ref.~\cite{Bergshoeff:1989de}, one is still dealing with a Supergravity
  theory, by definition. See, \textit{e.g.}~\cite{Ozkan:2024euj} for a
  review.} in $\alpha'$ stringy black-hole solutions is mostly the story of
the Supergravity black-hole solutions reviewed in Section~\ref{eq:susybhs}
coupled to a dictionary that translates the black-hole charges into brane
charges and relates their dualities (see, for instance,
Ref.~\cite{Balasubramanian:1997ak}). It was only natural to try to extend the
comparison between their macroscopic (BH) and microscopic entropies at zeroth
order in $\alpha'$ and $g_{s}$ to higher orders in both parameters.

\section{Beyond ``the Supergravity approximation''}

Since String Theory is expected to be a theory of Quantum Gravity, the
community expects it to solve the puzzles brought by the classical (GR) and
semiclassical descriptions of the interaction between gravity and matter that
we have, such as the information paradox, the endpoint of black-hole
evaporation, the meaning and value of the black-hole entropy, the unavoidable
presence of singularities in GR etc. Since, at lowest order in $\alpha'$ and
$g_{s}$ String Theory behaves as GR coupled to particular (but conventional)
kinds of matter, it is not surprising that, at this orders, it presents the
same problems, even if it provided a better understanding of the value of the
entropy of extremal black holes. The interest in studying the effect of higher
corrections in both parameters (stringy finite-size effects and stringy
quantum loop effects) on black-hole physics was driven by these expectations
as well as by the need to test the microscopic interpretation of the entropy.

Very little is known about $g_{s}$ corrections to the string effective action,
the exception being that of the IIB theory and of M~theory, thanks to duality
\cite{Green:1997tv}. Therefore, most of the work was focused on $\alpha'$
corrections, which are present at first order in the Heterotic Superstring
effective action and very well described up to third order as a higher-order
Supergravity by the Bergshoeff--de Roo action \cite{Bergshoeff:1989de} (a
remarkable improvement of the one given by Gross and Sloan in
Ref.~\cite{Gross:1986mw} in which supersymmetry is under control). Since
dealing with all the terms of higher order in the curvature made it very
difficult to calculate the corrections to complete black-hole solutions, some
researchers chose some particular terms that they considered relevant (see,
for instance, the pioneering work Ref.~\cite{Callan:1988hs} or the more recent
Ref.~\cite{Sahoo:2006pm}). In this way, Refs.~\cite{Dabholkar:2004yr}
presented results that hinted to the removal of the singularities of ``small''
black holes, although a closer analysis has disproven this claim
\cite{Cano:2018hut}.  A related approach made use of the corrections induced
in the symplectic formalism of $\mathcal{N}=2,d=4$ Supergravity
\cite{Behrndt:1998eq}.\footnote{For a review, see Ref.~\cite{Mohaupt:2000mj}.}
Others studied the corrections to the near-horizon geometry of the extremal
black holes and the attractor mechanism \cite{Castro:2007sd}. Sen's entropy
function formalism \cite{Sen:2005iz,Sen:2005wa} is based in this approach, see
Ref.~\cite{Sen:2007qy} for a review. This approach has also been successfully
used to compute logarithmic corrections to the black-hole entropy
\cite{Banerjee:2010qc}.

Since the entropy is a function of conserved charges which are defined
asymptotically,\footnote{In presence of $\alpha'$ corrections, most charges do
  not satisfy naive Gauss laws.} the values computed in the near-horizon
region may differ from the real ones. Furthermore, a given selection of
higher-order terms may not be a consistent truncation. Finding agreement with
the microscopic entropy does not guarantee that the value found for the
macroscopic entropy is correct. We want to compare two results whose validity
must be checked independently and, in the case of the macroscopic entropy,
this requires a calculation of the $\alpha'$ corrections to the complete black
hole geometry and to the entropy using all the terms in the effective action.

The first calculation of the first-order corrections in $\alpha'$ of the
complete geometry of a stringy black hole\footnote{The pioneering and almost
  forgotten calculation of Ref.~\cite{Campbell:1991kz} in which the $\alpha'$
  corrections to the 4-dimensional Kerr--Newman black holes were computed
  using a 4-dimensional version of the Gross--Sloan effective action of the
  Heterotic Superstring must also be mentioned here. The embedding of the
  electric charge in the Yang--Mills sector makes a comparison of its
  macroscopic entropy with the microscopic entropies computed elsewhere
  difficult. Furthermore, its non-rotating extremal limit is not
  supersymmetric.} (Strominger and Vafa's) using all the terms in the
Bergshoeff--de Roo action was published 22 years after that of the
zeroth-order solution in Ref.~\cite{Cano:2018qev} and those of the
Maldacena--Strominger--Johnson--Khuri--Myers followed suit
\cite{Cano:2018brq}.\footnote{The techniques used in those papers
  \cite{Chimento:2018kop} also showed how $\alpha'$ corrections could
  transform into a fully regular black hole the singular solutions known as
  \textit{massless black holes}, although this is a case in which the validity
  of the approximation is questionable \cite{Cano:2018aod}.}

Although it was initially claimed that the results for the macroscopic Wald)
entropy matched those found for the microscopic one (due to confirmation
bias), different results were obtained performing the calculation in different
dimensions \cite{Faedo:2019xii}.  It was soon realized that the ultimate cause
for these ambiguities was that the widely used Iyer--Wald prescription, which
does not account correctly for fields with gauge freedoms
\cite{Prabhu:2015vua}, fails to give a frame-independent entropy formula due
to the presence of Lorentz--Chern--Simons terms in the Kalb--Ramond field
strength. The correct treatment leads to a gauge-invariant formula
\cite{Elgood:2020nls} which gives a different $\alpha'$ correction to the
entropy. Since the defining property of the Wald entropy is that it satisfies

\begin{equation}
  \frac{\partial S}{\partial M}
  =
  \frac{1}{T}\,,  
\end{equation}

\noindent
its value can only be properly tested, avoiding confirmation biases, in
non-extremal black holes. This motivated the computation of the $\alpha'$
corrections of the non-extremal Reissner-Nordstr\"om
\cite{Cano:2019ycn},\footnote{Actually, of the particularly simple embedding
  with non-supersymmetric extremal limit \cite{Khuri:1995xq}.}  the
non-supersymmetric \cite{Cano:2021nzo} and non-extremal Strominger--Vafa
\cite{Cano:2022tmn}, the non-extremal
Maldacena--Strominger--Johnson--Khuri--Myers \cite{Zatti:2023oiq} and some
multicenter black-hole solutions \cite{Ortin:2021win}. In all these cases the
thermodynamic relation was satisfied by the entropy computed with the formula
proposed in Ref.~\cite{Elgood:2020nls}.

The thermodynamics of the $\alpha'$-corrected black holes contains surprises:
$\alpha'$ seems to play the role of a charge in the first law and in the Smarr
formula \cite{Zatti:2023oiq}, just as the components of the embedding tensor
do in gauged Supergravities \cite{Meessen:2022hcg}. Understanding this
behaviour is one of the many remaining challenges in this field. Other
important challenges have to do with the definition of the charges in presence
of $\alpha'$ corrections and their behaviour under duality, the treatment of
$\mathcal{O}(\alpha^{\prime\, 2})$ corrections etc.

\section{Supergravity black holes beyond 50?}

As this brief overview shows, the amount of work done in this field, specially
over the last 35 years, is huge. So is the number of important results
obtained, some of them revolutionary. Are there important questions yet to be
answered?\footnote{This is the polite form of asking: Has this field any
  future? Would you recommend a young researcher to enter this field?} The
span of attention of our society and also of the scientific community is
gradually decreasing.  Extending, confirming, disproving or correcting results
that, just a few years ago, attracted much attention and were considered very
important and even fundamental, may now be considered a futile effort. This
applies specially to the last two cases since, often, the community believes
that all the problems have been correctly solved and all the correct answers
have been obtained in its final form. However, the spirit of the scientific
enterprise is the pursuit of truth, questioning everything.  If the problems
we invested our lives trying to solve were truly important, perfecting the
solutions and reaching a deeper understanding must be considered equally
important, even if it is not fashionable.

Having said this, apart from the current research mentioned in some of the
previous sections, there are still many open or not completely solved problems
in this field: the explicit form of many black hole solutions (specially of
the non-extremal ones) are yet unknown; at first order in $\alpha'$, only a
few solutions and their Wald entropies have been computed (none at higher
orders); the explicit forms and properties of the thermodynamic (``chemical'')
potentials, which hold a much information and are assumed to be subject to
generalized zeroth laws have not been studied; exotic extremal black holes
with different fake central charges driving the scalars at infinity and the
horizon have not been studied etc.

From a more general perspective, black hole uniqueness,\footnote{In higher
  dimensions, one should include the topology of the horizon as a defining
  characteristic of black holes to recover uniqueness.} which is a keystone of
our understanding of black-hole physics, establishes a one-to-one relation
between theories and families of black-hole solutions\footnote{A very general
  question, which is somewhat related to chaos: is there a general form of the
  general family of black-hole solutions of a given theory valid for all the
  values of the charges and moduli? Are there examples in which small
  variations of the charges and moduli completely change the form of the
  solutions so that one has to solve the equations for each possible choice?}
and their entropies. Thus, it should be possible to compute them and other
properties without having to explicitly find the solutions. This is, to some
extent, the meaning of the attractor mechanism: for a given set of charges and
moduli, the theory determines the black-hole potential and, its extrema, the
entropy of extremal black holes, making it unnecessary to know the form of the
metric and other fields. Unfortunately, we do not know how to do the same in
the non-extremal case, but it is clear that Supergravity provides the best
background to attack this problem.

A healthy scientific community is a community with diverse researchers freely
working on diverse topics. Theoretical Physics should not be subject to
anybody's white papers nor to any form of \textit{pensamiento \'unico}. I
believe Supergravity will continue to be a source of important ideas, a
fertile playground for gravity and one of the main tools for research in
String Theory for many years to come.

\section*{Acknowledgments}

TO would like to thank all his collaborators over the years, senior and
junior, from which he has learned the little he knows and with whom he has
enjoyed working on Supergravity.  The work of TO has been supported in part by
the MCI, AEI, FEDER (UE) grants PID2021-125700NB-C21 (``Gravity, Supergravity
and Superstrings'' (GRASS)) and IFT Centro de Excelencia Severo Ochoa
CEX2020-001007-S. TO wishes to thank M.M.~Fern\'andez for her permanent
support.



\begin{thebibliography}{99}

\bibitem{kn:Schwarzschild}
K.~Schwarzschild,
Sitzungsber. deutsch. Akad.  Wiss. Berlin, Kl. Math. Phys. Technik (1916) 189--196.
English translation by S.~Antoci and A.~Loinger 
available in 
\href{http://arXiv.org/pdf/physics/9905030}{http://arXiv.org/pdf/physics/9905030}.
  J.~Droste,
Proc. Kon. Ned. Akad. Wet. \textbf{19 (1)} (1916) 197-215. 

\bibitem{kn:Reissner}
  H.~Reissner,
Ann. Phys. \textbf{50} (1916) 106.
G.~Nordstr\"om,
Proc. Kon. Ned. Akad. Wet. \textbf{20} (1918) 1238.

\bibitem{Tangherlini:1963bw}
F.~R.~Tangherlini,
Nuovo Cim. \textbf{27} (1963), 636-651

\bibitem{Kerr:1963ud}
R.~P.~Kerr,
Phys. Rev. Lett. \textbf{11} (1963), 237-238
E.~T.~Newman and A.~I.~Janis,
J. Math. Phys. \textbf{6} (1965), 915-917
E.~T.~Newman, R.~Couch, K.~Chinnapared, A.~Exton, A.~Prakash and R.~Torrence,
J. Math. Phys. \textbf{6} (1965), 918-919

\bibitem{Majumdar:1947eu}
S.~D.~Majumdar,
Phys. Rev. \textbf{72} (1947), 390-398
A.~Papapetrou,
Proc. Roy. Irish Acad. \textbf{A51} (1947) 191.
\href{https://www.jstor.org/stable/20488481}{JSTOR:20488481}

\bibitem{Hartle:1972ya}
J.~B.~Hartle and S.~W.~Hawking,
Commun. Math. Phys. \textbf{26} (1972), 87-101

\bibitem{Duff:1986hr}
M.~J.~Duff, B.~E.~W.~Nilsson and C.~N.~Pope,
Phys. Rept. \textbf{130} (1986), 1-142

\bibitem{Ortin:2015hya}
T.~Ort\'{\i}n,
Cambridge University Press, 2015.

\bibitem{Ferrara:1997tw}
S.~Ferrara, G.~W.~Gibbons and R.~Kallosh,
Nucl. Phys. B \textbf{500} (1997), 75-93
[\hepth{9702103} [hep-th]

\bibitem{DAuria:1998emc}
R.~D'Auria and P.~Fr\'e,
Contribution to the  8th Graduate School in Contempory Relativity and
Gravitational Physics: The Physics of Black Holes (SIGRAV 98), 137-272
[\hepth{9812160} [hep-th]].

\bibitem{Mohaupt:2000mj}
T.~Mohaupt,
Fortsch. Phys. \textbf{49} (2001), 3-161
[\hepth{0007195} [hep-th]].

\bibitem{Andrianopoli:2006ub}
L.~Andrianopoli, R.~D'Auria, S.~Ferrara and M.~Trigiante,
Lect. Notes Phys. \textbf{737} (2008), 661-727
[\hepth{0611345} [hep-th]].

\bibitem{Ferrara:2008hwa}
S.~Ferrara, K.~Hayakawa and A.~Marrani,
Fortsch. Phys. \textbf{56} (2008) no.10, 993-1046
[\arxiv{0805.2498} [hep-th]].

\bibitem{Bellucci:2009qv}
S.~Bellucci, S.~Ferrara, M.~Gunaydin and A.~Marrani,
Springer Proc. Phys. \textbf{134} (2010), 1-30
[\arxiv{0905.3739} [hep-th]].

\bibitem{Ortin:2016xvk}
T.~Ort\'{\i}n and P.~F.~Ram\'{\i}rez,
Springer Proc. Phys. \textbf{176} (2016), 1-40

\bibitem{Baaklini:1977bx}
N.~S.~Baaklini, S.~Ferrara and P.~van Nieuwenhuizen,
Lett. Nuovo Cim. \textbf{20} (1977), 113

\bibitem{Cordero:1978ud}
P.~Cordero and C.~Teitelboim,
Phys. Lett. B \textbf{78} (1978), 80-83

\bibitem{Gueven:1980be}
R.~Gueven,
Phys. Rev. D \textbf{22} (1980), 2327

\bibitem{Aichelburg:1981rd}
P.~C.~Aichelburg and R.~Gueven,
Phys. Rev. D \textbf{24} (1981), 2066

\bibitem{Gueven:1982tk}
R.~Gueven,
Phys. Rev. D \textbf{25} (1982), 3117

\bibitem{Aichelburg:1983ux}
P.~C.~Aichelburg and R.~Gueven,
Phys. Rev. Lett. \textbf{51} (1983), 1613

\bibitem{kn:Embacher1984}
  F.~Embacher,
Gen. Rel. Grav. \textbf{16} (1984) 909.
F.~Embacher and P.~C.~Aichelburg,
Phys. Rev. D \textbf{30} (1984), 2457-2461
P.~C.~Aichelburg and F.~Embacher,
Phys. Rev. D \textbf{34} (1986), 3006
\doi{10.1103/PhysRevD.34.3006}

\bibitem{Gibbons:1982ih}
G.~W.~Gibbons,
Nucl. Phys. B \textbf{207} (1982), 337-349

\bibitem{Gibbons:1987ps}
G.~W.~Gibbons and K.~i.~Maeda,
Nucl. Phys. B \textbf{298} (1988), 741-775

\bibitem{Garfinkle:1990qj}
D.~Garfinkle, G.~T.~Horowitz and A.~Strominger,
Phys. Rev. D \textbf{43} (1991), 3140
[erratum: Phys. Rev. D \textbf{45} (1992), 3888]

\bibitem{Kallosh:1992ii}
R.~Kallosh, A.~D.~Linde, T.~Ort\'{\i}n, A.~W.~Peet and A.~Van Proeyen,
Phys. Rev. D \textbf{46} (1992), 5278-5302
[\hepth{9205027} [hep-th]].

\bibitem{Witten:1981mf}
E.~Witten,
Commun. Math. Phys. \textbf{80} (1981), 381.
J.~A.~Nester,
Phys. Lett. A \textbf{83} (1981), 241.
W.~Israel and J.~A.~Nester,
Phys. Lett. A \textbf{85} (1981), 259,

\bibitem{Hull:1983ap}
C.~M.~Hull,
Commun. Math. Phys. \textbf{90} (1983), 545

\bibitem{Deser:1977hu}
S.~Deser and C.~Teitelboim,
Phys. Rev. Lett. \textbf{39} (1977), 249.
M.~T.~Grisaru,
Phys. Lett. B \textbf{73} (1978), 207-208.

\bibitem{Bogomolny:1975de}
E.~B.~Bogomolny,
Sov. J. Nucl. Phys. \textbf{24} (1976), 449.
%
M.~K.~Prasad and C.~M.~Sommerfield,
Phys. Rev. Lett. \textbf{35} (1975), 760-762.

\bibitem{Gibbons:1982fy}
G.~W.~Gibbons and C.~M.~Hull,
Phys. Lett. B \textbf{109} (1982), 190-194

\bibitem{Figueroa-OFarrill:1999klq}
J.~M.~Figueroa-O'Farrill,
Class. Quant. Grav. \textbf{16} (1999), 2043-2055
[\hepth{9902066} [hep-th]].
M.~Blau, J.~M.~Figueroa-O'Farrill, C.~Hull and G.~Papadopoulos,
JHEP \textbf{01} (2002), 047
[\hepth{0110242} [hep-th]].

\bibitem{Gibbons:1984kp}
G.~W.~Gibbons,
Print-85-0061 (CAMBRIDGE).
Contribution to the XV GIFT Seminar on Supersymmetry and Supergravity

\bibitem{Kallosh:1993wx}
R.~Kallosh and T.~Ort\'{\i}n,
[\hepth{9306085} [hep-th]].

\bibitem{Bellorin:2005hy}
J.~Bellor\'{\i}n and T.~Ort\'{\i}n,
Phys. Lett. B \textbf{616} (2005), 118-124
[\hepth{0501246} [hep-th]].

\bibitem{Scherk:1979aj}
J.~Scherk,
Phys. Lett. B \textbf{88} (1979), 265-267

\bibitem{Bellorin:2006xr}
J.~Bellor\'{\i}n, P.~Meessen and T.~Ort\'{\i}n,
Nucl. Phys. B \textbf{762} (2007), 229-255
[\hepth{0606201} [hep-th]].

\bibitem{Gauntlett:1998fz}
J.~P.~Gauntlett, R.~C.~Myers and P.~K.~Townsend,
Class. Quant. Grav. \textbf{16} (1999), 1-21
[\hepth{9810204} [hep-th]].

\bibitem{Shapere:1991ta}
A.~D.~Shapere, S.~Trivedi and F.~Wilczek,
Mod. Phys. Lett. A \textbf{6} (1991), 2677-2686

\bibitem{Ortin:1992ur}
T.~Ort\'{\i}n,
Phys. Rev. D \textbf{47} (1993), 3136-3143
[\hepth{9208078} [hep-th]].

\bibitem{Bergshoeff:1996gg}
E.~Bergshoeff, R.~Kallosh and T.~Ort\'{\i}n,
Nucl. Phys. B \textbf{478} (1996), 156-180
[\hepth{9605059} [hep-th]].

\bibitem{Lozano-Tellechea:1999lwm}
E.~Lozano-Tellechea and T.~Ort\'{\i}n,
black hole solutions of N=4, D = 4 supergravity,''
Nucl. Phys. B \textbf{569} (2000), 435-450
[\hepth{9910020} [hep-th]].

\bibitem{Sen:1994eb}
A.~Sen,
Nucl. Phys. B \textbf{440} (1995), 421-440
[\hepth{9411187} [hep-th]].
M.~Cveti\v{c} and D.~Youm,
Nucl. Phys. B \textbf{472} (1996), 249-267
[\hepth{9512127} [hep-th]].

\bibitem{Cvetic:1996zq}
M.~Cveti\v{c} and C.~M.~Hull,
Nucl. Phys. B \textbf{480} (1996), 296-316
[\hepth{9606193} [hep-th]].

\bibitem{Bertolini:1999je}
M.~Bertolini, P.~Fr\'e and M.~Trigiante,
Class. Quant. Grav. \textbf{16} (1999), 2987-3004
[\hepth{9905143} [hep-th]].

\bibitem{Ortin:2012gg}
T.~Ort\'{\i}n and C.~S.~Shahbazi,
Phys. Rev. D \textbf{86} (2012), 061702
[\arxiv{1206.3190} [hep-th]].

\bibitem{Tod:1983pm}
K.~P.~Tod,
Phys. Lett. B \textbf{121} (1983), 241-244

\bibitem{Perjes:1971gv}
Z.~Perj\'es,
Phys. Rev. Lett. \textbf{27} (1971), 1668. 
W.~Israel and G.~A.~Wilson,
J. Math. Phys. \textbf{13} (1972), 865-871

\bibitem{Denef:2000nb}
F.~Denef,
JHEP \textbf{08} (2000), 050
[\hepth{0005049} [hep-th]].
B.~Bates and F.~Denef,
JHEP \textbf{11} (2011), 127
[\hepth{0304094} [hep-th]].

\bibitem{Meessen:2017rwm}
P.~Meessen, T.~Ort\'{\i}n and P.~F.~Ram\'{\i}rez,
JHEP \textbf{10} (2017), 066
[\arxiv{1707.03846} [hep-th]].

\bibitem{Tod:1995jf}
K.~P.~Tod,
Class. Quant. Grav. \textbf{12} (1995), 1801-1820

\bibitem{Gauntlett:2002nw}
J.~P.~Gauntlett, J.~B.~Gutowski, C.~M.~Hull, S.~Pakis and H.~S.~Reall,
Class. Quant. Grav. \textbf{20} (2003), 4587-4634
[\hepth{0209114} [hep-th]].

\bibitem{Meessen:2001vx}
P.~Meessen,
Phys. Rev. D \textbf{65} (2002), 087501
[\hepth{0111031} [hep-th]].

\bibitem{kn:Penrose1976}
  R.~Penrose, in \textit{Differential Geometry and Relativity},
  Reidel, Dordrecht (1976) 271-275.
R.~Gueven,
Phys. Lett. B \textbf{482} (2000), 255-263
[\hepth{0005061} [hep-th]].
M.~Blau, J.~M.~Figueroa-O'Farrill, C.~Hull and G.~Papadopoulos,
Class. Quant. Grav. \textbf{19} (2002), L87-L95
[\hepth{0201081} [hep-th]].

\bibitem{Breckenridge:1996is}
J.~C.~Breckenridge, R.~C.~Myers, A.~W.~Peet and C.~Vafa,
Phys. Lett. B \textbf{391} (1997), 93-98
[\hepth{9602065} [hep-th]].

\bibitem{Elvang:2004rt}
H.~Elvang, R.~Emparan, D.~Mateos and H.~S.~Reall,
Phys. Rev. Lett. \textbf{93} (2004), 211302
[\hepth{0407065} [hep-th]].

\bibitem{Gauntlett:2004qy}
J.~P.~Gauntlett and J.~B.~Gutowski,
Phys. Rev. D \textbf{71} (2005), 045002
[\hepth{0408122} [hep-th]].

\bibitem{Ortin:2004af}
T.~Ort\'{\i}n,
Class. Quant. Grav. \textbf{22} (2005), 939-946
[\hepth{0410252} [hep-th]].

\bibitem{Kunduri:2014kja}
H.~K.~Kunduri and J.~Lucietti,
Phys. Rev. Lett. \textbf{113} (2014) no.21, 211101
[\arxiv{1408.6083} [hep-th]].
H.~K.~Kunduri and J.~Lucietti,
Phys. Rev. D \textbf{94} (2016) no.6, 064007
[\arxiv{1605.01545} [hep-th]].

\bibitem{Kunduri:2013gce}
H.~K.~Kunduri and J.~Lucietti,
Living Rev. Rel. \textbf{16} (2013), 8
[\arxiv{1306.2517} [hep-th]].

\bibitem{Gauntlett:2003fk}
J.~P.~Gauntlett and J.~B.~Gutowski,
Phys. Rev. D \textbf{68} (2003), 105009
[erratum: Phys. Rev. D \textbf{70} (2004), 089901]
[\hepth{0304064} [hep-th]].

\bibitem{Gutowski:2004yv}
J.~B.~Gutowski and H.~S.~Reall,
JHEP \textbf{04} (2004), 048
[\hepth{0401129} [hep-th]].
J.~B.~Gutowski and W.~Sabra,
JHEP \textbf{10} (2005), 039
\doi{10.1088/1126-6708/2005/10/039}
[\hepth{0505185} [hep-th]].

\bibitem{Bellorin:2006yr}
J.~Bellor\'{\i}n, P.~Meessen and T.~Ort\'{\i}n,
JHEP \textbf{01} (2007), 020
[\hepth{0610196} [hep-th]].
J.~Bellor\'{\i}n and T.~Ort\'{\i}n,
JHEP \textbf{08} (2007), 096
[\arxiv{0705.2567} [hep-th]].
J.~Bellor\'{\i}n,
Class. Quant. Grav. \textbf{26} (2009), 195012
[\arxiv{0810.0527} [hep-th]].

\bibitem{Coleman:1991ku}
S.~R.~Coleman, J.~Preskill and F.~Wilczek,
Nucl. Phys. B \textbf{378} (1992), 175-246
[\hepth{9201059} [hep-th]].

\bibitem{Meessen:2015enl}
P.~Meessen, T.~Ort\'{\i}n and P.~F.~Ram\'{\i}rez,
JHEP \textbf{03} (2016), 112
[\arxiv{1512.07131} [hep-th]].

\bibitem{Caldarelli:2003pb}
M.~M.~Caldarelli and D.~Klemm,
JHEP \textbf{09} (2003), 019
[\hepth{0307022} [hep-th]].

\bibitem{Meessen:2006tu}
P.~Meessen and T.~Ort\'{\i}in,
supergravity coupled to vector supermultiplets,''
Nucl. Phys. B \textbf{749} (2006), 291-324
[\hepth{0603099} [hep-th]].

\bibitem{Behrndt:1997ny}
K.~Behrndt, D.~L\"ust and W.~A.~Sabra,
Nucl. Phys. B \textbf{510} (1998), 264-288
[\hepth{9705169} [hep-th]].

\bibitem{Huebscher:2006mr}
M.~Huebscher, P.~Meessen and T.~Ort\'{\i}n,
Nucl. Phys. B \textbf{759} (2006), 228-248
[\hepth{0606281} [hep-th]].

\bibitem{Cacciatori:2008ek}
S.~L.~Cacciatori, D.~Klemm, D.~S.~Mansi and E.~Zorzan,
JHEP \textbf{05} (2008), 097
[\arxiv{0804.0009} [hep-th]].
D.~Klemm and E.~Zorzan,
Class. Quant. Grav. \textbf{26} (2009), 145018
[\arxiv{0902.4186} [hep-th]].

\bibitem{Huebscher:2007hj}
M.~H\"ubscher, P.~Meessen, T.~Ort\'{\i}n and S.~Vaul\`a,
Phys. Rev. D \textbf{78} (2008), 065031
[\arxiv{0712.1530} [hep-th]].
M.~H\"bscher, P.~Meessen, T.~Ort\'{\i}n and S.~Vaul\'a,
JHEP \textbf{09} (2008), 099
[\arxiv{0806.1477} [hep-th]].

\bibitem{Meessen:2012sr}
P.~Meessen and T.~Ort\'{\i}n,
Nucl. Phys. B \textbf{863} (2012), 65-89
[\arxiv{1204.0493} [hep-th]].

\bibitem{Meessen:2008kb}
P.~Meessen,
Phys. Lett. B \textbf{665} (2008), 388-391
[\arxiv{0803.0684} [hep-th]].

\bibitem{Bellorin:2005zc}
J.~Bellor\'{\i}n and T.~Ort\'{\i}n,
Nucl. Phys. B \textbf{726} (2005), 171-209
[\hepth{0506056} [hep-th]].

\bibitem{Meessen:2010fh}
P.~Meessen, T.~Ort\'{\i}n and S.~Vaul\`a,
JHEP \textbf{11} (2010), 072
[\arxiv{1006.0239} [hep-th]].

\bibitem{Shmakova:1996nz}
M.~Shmakova,
Phys. Rev. D \textbf{56} (1997), 540-544
[\hepth{9612076} [hep-th]].

\bibitem{Gutowski:2003rg}
J.~B.~Gutowski, D.~Martelli and H.~S.~Reall,
Class. Quant. Grav. \textbf{20} (2003), 5049-5078
[\hepth{0306235} [hep-th]].

\bibitem{Cariglia:2004kk}
M.~Cariglia and O.~A.~P.~Mac Conamhna,
Class. Quant. Grav. \textbf{21} (2004), 3171-3196
[\hepth{0402055} [hep-th]].
M.~Akyol and G.~Papadopoulos,
Class. Quant. Grav. \textbf{28} (2011), 105001
[\arxiv{1010.2632} [hep-th]].
H.~Het Lam and S.~Vandoren,
JHEP \textbf{06} (2018), 021
[\arxiv{1804.04681} [hep-th]].
P.~A.~Cano and T.~Ort\'{\i}n,
Class. Quant. Grav. \textbf{36} (2019) no.12, 125007
[\arxiv{1804.04945} [hep-th]].

\bibitem{Khuri:1995xq}
R.~R.~Khuri and T.~Ort\'{\i}n,
Phys. Lett. B \textbf{373} (1996), 56-60
[\hepth{9512178} [hep-th]].

\bibitem{Ortin:1996bz}
T.~Ort\'{\i}n,
Phys. Lett. B \textbf{422} (1998), 93-100
[\hepth{9612142} [hep-th]].

\bibitem{Ceresole:2007wx}
A.~Ceresole and G.~Dall'Agata,
JHEP \textbf{03} (2007), 110
[\hepth{0702088} [hep-th]].

\bibitem{Mohaupt:2009iq}
T.~Mohaupt and K.~Waite,
JHEP \textbf{10} (2009), 058
[\arxiv{0906.3451} [hep-th]].
T.~Mohaupt and O.~Vaughan,
JHEP \textbf{07} (2012), 163
[\arxiv{1112.2876} [hep-th]].

\bibitem{Galli:2011fq}
P.~Galli, T.~Ort\'{\i}n, J.~Perz and C.~S.~Shahbazi,
JHEP \textbf{07} (2011), 041
[\arxiv{1105.3311} [hep-th]].
P.~Meessen and T.~Ort\'{\i}n,
Phys. Lett. B \textbf{707} (2012), 178-183
[\arxiv{1107.5454} [hep-th]].

\bibitem{Meessen:2011aa}
P.~Meessen, T.~Ort\'{\i}n, J.~Perz and C.~S.~Shahbazi,
Phys. Lett. B \textbf{709} (2012), 260-265
[\arxiv{1112.3332} [hep-th]].
P.~Meessen, T.~Ort\'{\i}n, J.~Perz and C.~S.~Shahbazi,
JHEP \textbf{09} (2012), 001
[\arxiv{1204.0507} [hep-th]].
P.~Galli, T.~Ort\'{\i}n, J.~Perz and C.~S.~Shahbazi,
JHEP \textbf{04} (2013), 157
\doi{10.1007/JHEP04(2013)157}
[\arxiv{1212.0303} [hep-th]].

\bibitem{Ferrara:1995ih}
S.~Ferrara, R.~Kallosh and A.~Strominger,
Phys. Rev. D \textbf{52} (1995), R5412-R5416
[\hepth{9508072} [hep-th]].
A.~Strominger,
Phys. Lett. B \textbf{383} (1996), 39-43
[\hepth{9602111} [hep-th]].
S.~Ferrara and R.~Kallosh,
Phys. Rev. D \textbf{54} (1996), 1514-1524
[\hepth{9602136} [hep-th]].
S.~Ferrara and R.~Kallosh,
Phys. Rev. D \textbf{54} (1996), 1525-1534
[\hepth{9603090} [hep-th]].

\bibitem{deAntonioMartin:2012bi}
A.~de Antonio Mart\'{\i}n, T.~Ort\'{\i}n and C.~S.~Shahbazi,
JHEP \textbf{05} (2012), 045
[\arxiv{1203.0260} [hep-th]].

\bibitem{Pioline:2006ni}
B.~Pioline,
Class. Quant. Grav. \textbf{23} (2006), S981
[\hepth{0607227} [hep-th]].
B.~Pioline,
Lect. Notes Phys. \textbf{755} (2008), 283-373

\bibitem{Breitenlohner:1987dg}
P.~Breitenlohner, D.~Maison and G.~W.~Gibbons,
Commun. Math. Phys. \textbf{120} (1988), 295

\bibitem{Borsten:2009zy}
L.~Borsten, D.~Dahanayake, M.~J.~Duff and W.~Rubens,
Phys. Rev. D \textbf{80} (2009), 026003
[\arxiv{0903.5517} [hep-th]].

\bibitem{Ferrara:2011gv}
S.~Ferrara, A.~Marrani and A.~Yeranyan,
Phys. Lett. B \textbf{701} (2011), 640-645
[\arxiv{1102.4857} [hep-th]].

\bibitem{Galli:2012ji}
P.~Galli, P.~Meessen and T.~Ort\'{\i}n,
JHEP \textbf{05} (2013), 011
[\arxiv{1211.7296} [hep-th]].

\bibitem{Ballesteros:2023iqb}
R.~Ballesteros, C.~G\'omez-Fayr\'en, T.~Ort\'{\i}n and M.~Zatti,
JHEP \textbf{05} (2023), 158
[\arxiv{2302.11630} [hep-th]].

\bibitem{Gibbons:1996af}
G.~W.~Gibbons, R.~Kallosh and B.~Kol,
Phys. Rev. Lett. \textbf{77} (1996), 4992-4995
[\hepth{9607108} [hep-th]].

\bibitem{Sen:2005iz}
A.~Sen,
JHEP \textbf{03} (2006), 008
[\hepth{0508042} [hep-th]].].

\bibitem{Andrianopoli:1996cm}
L.~Andrianopoli, M.~Bertolini, A.~Ceresole, R.~D'Auria,
S.~Ferrara, P.~Fr\'e and T.~Magri,
J. Geom. Phys. \textbf{23} (1997), 111-189
[\hepth{9605032} [hep-th]].

\bibitem{Goldstein:2005hq}
K.~Goldstein, N.~Iizuka, R.~P.~Jena and S.~P.~Trivedi,
Phys. Rev. D \textbf{72} (2005), 124021
[\hepth{0507096} [hep-th]].
P.~K.~Tripathy and S.~P.~Trivedi,
JHEP \textbf{03} (2006), 022
[\hepth{0511117} [hep-th]].

\bibitem{LopesCardoso:2007qid}
G.~Lopes Cardoso, A.~Ceresole, G.~Dall'Agata,
J.~M.~Oberreuter and J.~Perz,
JHEP \textbf{10} (2007), 063
[\arxiv{0706.3373} [hep-th]].

\bibitem{Miller:2006ay}
C.~M.~Miller, K.~Schalm and E.~J.~Weinberg,
Phys. Rev. D \textbf{76} (2007), 044001
[\hepth{0612308} [hep-th]].
D.~Klemm and M.~Rabbiosi,
JHEP \textbf{10} (2017), 149
[\arxiv{1706.05862} [hep-th]].

\bibitem{Skenderis:2006rr}
K.~Skenderis and P.~K.~Townsend,
Phys. Rev. D \textbf{74} (2006), 125008
[\hepth{0609056} [hep-th]].
M.~Trigiante, T.~Van Riet and B.~Vercnocke,
JHEP \textbf{05} (2012), 078
[\arxiv{1203.3194} [hep-th]].

\bibitem{Bergshoeff:1989de}
E.~A.~Bergshoeff and M.~de Roo,
Nucl. Phys. B \textbf{328} (1989), 439-468

\bibitem{Gaillard:1981rj}
M.~K.~Gaillard and B.~Zumino,
Nucl. Phys. B \textbf{193} (1981), 221-244

\bibitem{Papadopoulos:1996uq}
G.~Papadopoulos, P.~K.~Townsend and P.~V.~Landshoff,
Phys. Lett. B \textbf{380} (1996), 273-279
[\hepth{9603087} [hep-th]].

\bibitem{Duff:1994an}
M.~J.~Duff, R.~R.~Khuri and J.~X.~Lu,
Phys. Rept. \textbf{259} (1995), 213-326
[\hepth{9412184} [hep-th]].

\bibitem{Witten:1995ex}
E.~Witten,
Nucl. Phys. B \textbf{443} (1995), 85-126
[\hepth{9503124} [hep-th]].

\bibitem{Polchinski:1995mt}
J.~Polchinski,
Phys. Rev. Lett. \textbf{75} (1995), 4724-4727
\doi{10.1103/PhysRevLett.75.4724}
[\hepth{9510017} [hep-th]].

\bibitem{Bergshoeff:1995as}
E.~Bergshoeff, C.~M.~Hull and T.~Ort\'{\i}n,
Nucl. Phys. B \textbf{451} (1995), 547-578
[\hepth{9504081} [hep-th]].

\bibitem{Hull:1994ys}
C.~M.~Hull and P.~K.~Townsend,
Nucl. Phys. B \textbf{438} (1995), 109-137
[\hepth{9410167} [hep-th]].

\bibitem{Cano:2021rey}
P.~A.~Cano and A.~Ruip\'erez,
Phys. Rev. D \textbf{105} (2022) no.4, 044022
[\arxiv{2111.04750} [hep-th]].

\bibitem{Cano:2018qev}
P.~A.~Cano, P.~Meessen, T.~Ort\'{\i}n and P.~F.~Ram\'{\i}rez,
JHEP \textbf{05} (2018), 110
[\arxiv{1803.01919} [hep-th]].

\bibitem{Cano:2018brq}
  P.~A.~Cano, S.~Chimento, P.~Meessen, T.~Ort\'{\i}n,
  P.~F.~Ram\'{\i}rez and A.~Ruip\'erez,
JHEP \textbf{02} (2019), 192
[\arxiv{1808.03651} [hep-th]].

\bibitem{Duff:1995sm}
M.~J.~Duff, J.~T.~Liu and J.~Rahmfeld,
Nucl. Phys. B \textbf{459} (1996), 125-159
[\hepth{9508094} [hep-th]].
R.~R.~Khuri and T.~Ort\'{\i}n,
Nucl. Phys. B \textbf{467} (1996), 355-382
[\hepth{9512177} [hep-th]].

\bibitem{Maldacena:1996gb}
J.~M.~Maldacena and A.~Strominger,
Phys. Rev. Lett. \textbf{77} (1996), 428-429
[\hepth{9603060} [hep-th]].
C.~V.~Johnson, R.~R.~Khuri and R.~C.~Myers,
Phys. Lett. B \textbf{378} (1996), 78-86
[\hepth{9603061} [hep-th]].

\bibitem{Strominger:1996sh}
A.~Strominger and C.~Vafa,
Phys. Lett. B \textbf{379} (1996), 99-104
[\hepth{9601029} [hep-th]].

\bibitem{Maldacena:1997de}
J.~M.~Maldacena, A.~Strominger and E.~Witten,
JHEP \textbf{12} (1997), 002
[\hepth{9711053} [hep-th]].

\bibitem{Gibbons:1994ff}
G.~W.~Gibbons and R.~E.~Kallosh,
Phys. Rev. D \textbf{51} (1995), 2839-2862
[\hepth{9407118} [hep-th]].
S.~W.~Hawking, G.~T.~Horowitz and S.~F.~Ross,
Phys. Rev. D \textbf{51} (1995), 4302-4314
[\grqc{9409013} [gr-qc]].

\bibitem{Kallosh:1992wa}
R.~Kallosh, T.~Ort\'{\i}n and A.~W.~Peet,
Phys. Rev. D \textbf{47} (1993), 5400-5407
[\hepth{9211015} [hep-th]].

\bibitem{Iliesiu:2020qvm}
L.~V.~Iliesiu and G.~J.~Turiaci,
JHEP \textbf{05} (2021), 145
[\arxiv{2003.02860} [hep-th]].
M.~Heydeman, L.~V.~Iliesiu, G.~J.~Turiaci and W.~Zhao,
J. Phys. A \textbf{55} (2022) no.1, 014004
[\arxiv{2011.01953} [hep-th]].
M.~Kolanowski, D.~Marolf, I.~Rakic, M.~Rangamani and G.~J.~Turiaci,
[\arxiv{2409.16248} [hep-th]].

\bibitem{Ozkan:2024euj}
M.~Ozkan, Y.~Pang and E.~Sezgin,
Phys. Rept. \textbf{1086} (2024), 1-95
[\arxiv{2401.08945} [hep-th]].

\bibitem{Balasubramanian:1997ak}
V.~Balasubramanian,
NATO Sci. Ser. C \textbf{520} (1999), 399-410
[\hepth{9712215} [hep-th]].

\bibitem{Green:1997tv}
M.~B.~Green and M.~Gutperle,
Nucl. Phys. B \textbf{498} (1997), 195-227
[\hepth{9701093} [hep-th]].
M.~B.~Green and P.~Vanhove,
Phys. Lett. B \textbf{408} (1997), 122-134
[\hepth{9704145} [hep-th]].
M.~B.~Green, M.~Gutperle and P.~Vanhove,
Phys. Lett. B \textbf{409} (1997), 177-184
[\hepth{9706175} [hep-th]].

\bibitem{Gross:1986mw}
D.~J.~Gross and J.~H.~Sloan,
Nucl. Phys. B \textbf{291} (1987), 41-89

\bibitem{Callan:1988hs}
C.~G.~Callan, Jr., R.~C.~Myers and M.~J.~Perry,
Nucl. Phys. B \textbf{311} (1989), 673-698

\bibitem{Sahoo:2006pm}
B.~Sahoo and A.~Sen,
JHEP \textbf{01} (2007), 010
[\hepth{0608182} [hep-th]].

\bibitem{Dabholkar:2004yr}
A.~Dabholkar,
Phys. Rev. Lett. \textbf{94} (2005), 241301
[\hepth{0409148} [hep-th]].
A.~Dabholkar, R.~Kallosh and A.~Maloney,
JHEP \textbf{12} (2004), 059
[\hepth{0410076} [hep-th]].

\bibitem{Cano:2018hut}
P.~A.~Cano, P.~F.~Ram\'{\i}rez and A.~Ruip\'erez,
JHEP \textbf{03} (2020), 115
[\arxiv{1808.10449} [hep-th]].
A.~Ruip\'erez,
Class. Quant. Grav. \textbf{38} (2021) no.14, 145011
[\arxiv{2003.02269} [hep-th]].
P.~A.~Cano, \'A.~Murcia, P.~F.~Ram\'{\i}rez and A.~Ruip\'erez,
JHEP \textbf{05} (2021), 272
[\arxiv{2102.04476} [hep-th]].
Y.~Chen, J.~Maldacena and E.~Witten,
JHEP \textbf{01} (2023), 103
[\arxiv{2109.08563} [hep-th]].
S.~Massai, A.~Ruip\'erez and M.~Zatti,
JHEP \textbf{04} (2024), 150
[\arxiv{2311.03308} [hep-th]].

\bibitem{Behrndt:1998eq}
K.~Behrndt, G.~Lopes Cardoso, B.~de Wit, D.~L\"ust,
T.~Mohaupt and W.~A.~Sabra,
Phys. Lett. B \textbf{429} (1998), 289-296
[\hepth{9801081} [hep-th]].
G.~Lopes Cardoso, B.~de Wit and T.~Mohaupt,
Phys. Lett. B \textbf{451} (1999), 309-316
[\hepth{9812082} [hep-th]].
G.~Lopes Cardoso, B.~de Wit and T.~Mohaupt,
Fortsch. Phys. \textbf{48} (2000), 49-64
[\hepth{9904005} [hep-th]].
G.~Lopes Cardoso, B.~de Wit and T.~Mohaupt,
Class. Quant. Grav. \textbf{17} (2000), 1007-1015
[\hepth{9910179} [hep-th]].
G.~Lopes Cardoso, B.~de Wit, J.~Kappeli and T.~Mohaupt,
JHEP \textbf{12} (2000), 019
[\hepth{0009234} [hep-th]].
G.~Lopes Cardoso, B.~de Wit, J.~Kappeli and T.~Mohaupt,
theories with R**2 interactions,''
Fortsch. Phys. \textbf{49} (2001), 557-563
[\hepth{0012232} [hep-th]].

\bibitem{Castro:2007sd}
A.~Castro, J.~L.~Davis, P.~Kraus and F.~Larsen,
JHEP \textbf{04} (2007), 091
[\hepth{0702072} [hep-th]].
A.~Castro, J.~L.~Davis, P.~Kraus and F.~Larsen,
JHEP \textbf{06} (2007), 007
[\hepth{0703087} [hep-th]].
A.~Castro, J.~L.~Davis, P.~Kraus and F.~Larsen,
Int. J. Mod. Phys. A \textbf{23} (2008), 613-691
[\arxiv{0801.1863} [hep-th]].

\bibitem{Sen:2005wa}
A.~Sen,
JHEP \textbf{09} (2005), 038
[\hepth{0506177} [hep-th]].

\bibitem{Sen:2007qy}
A.~Sen,
Gen. Rel. Grav. \textbf{40} (2008), 2249-2431
[\arxiv{0708.1270} [hep-th]].

\bibitem{Banerjee:2010qc}
S.~Banerjee, R.~K.~Gupta and A.~Sen,
JHEP \textbf{03} (2011), 147
[\arxiv{1005.3044} [hep-th]].
S.~Banerjee, R.~K.~Gupta, I.~Mandal and A.~Sen,
JHEP \textbf{11} (2011), 143
[\arxiv{1106.0080} [hep-th]].
A.~Sen,
Gen. Rel. Grav. \textbf{44} (2012) no.5, 1207-1266
[\arxiv{1108.3842} [hep-th]].
A.~Sen,
Gen. Rel. Grav. \textbf{44} (2012), 1947-1991
[\arxiv{1109.3706} [hep-th]].
A.~Sen,
JHEP \textbf{04} (2013), 156
[\arxiv{1205.0971} [hep-th]].

\bibitem{Campbell:1991kz}
B.~A.~Campbell, N.~Kaloper and K.~A.~Olive,
Phys. Lett. B \textbf{285} (1992), 199-205

\bibitem{Chimento:2018kop}
  S.~Chimento, P.~Meessen, T.~Ort\'{\i}n,
  P.~F.~Ram\'{\i}rez and A.~Ruip\'erez,
JHEP \textbf{07} (2018), 080
[\arxiv{1803.04463} [hep-th]].

\bibitem{Cano:2018aod}
P.~A.~Cano, S.~Chimento, T.~Ort\'{\i}n and A.~Ruip\'erez,
Phys. Rev. D \textbf{99} (2019) no.4, 046014
[\arxiv{1806.08377} [hep-th]].


\bibitem{Faedo:2019xii}
F.~Faedo and P.~F.~Ram\'{\i}rez,
JHEP \textbf{10} (2019), 033
[\arxiv{1906.12287} [hep-th]].

\bibitem{Prabhu:2015vua}
K.~Prabhu,
Class. Quant. Grav. \textbf{34} (2017) no.3, 035011
[\arxiv{1511.00388} [gr-qc]].

\bibitem{Elgood:2020nls}
Z.~Elgood, T.~Ort\'{\i}n and D.~Pere\~niguez,
JHEP \textbf{05} (2021), 110
[\arxiv{2012.14892} [hep-th]].

\bibitem{Cano:2019ycn}
P.~A.~Cano, S.~Chimento, R.~Linares, T.~Ort\'{\i}n and P.~F.~Ram\'{\i}rez,
JHEP \textbf{02} (2020), 031
[\arxiv{1910.14324} [hep-th]].

\bibitem{Cano:2021nzo}
P.~A.~Cano, T.~Ort\'{\i}n, A.~Ruip\'erez and M.~Zatti,
JHEP \textbf{03} (2022), 103
[\arxiv{2111.15579} [hep-th]].

\bibitem{Cano:2022tmn}
P.~A.~Cano, T.~Ort\'{\i}n, A.~Ruip\'erez and M.~Zatti,
JHEP \textbf{12} (2022), 150
[\arxiv{2210.01861} [hep-th]].

\bibitem{Zatti:2023oiq}
M.~Zatti,
4-dimensional non-extremal stringy black holes,''
JHEP \textbf{11} (2023), 185
[\arxiv{2308.12879} [hep-th]].

\bibitem{Ortin:2021win}
T.~Ort\'{\i}n, A.~Ruip\'erez and M.~Zatti,
SciPost Phys. Core \textbf{6} (2023) no.4, 072
[\arxiv{2112.12764} [hep-th]].

\bibitem{Meessen:2022hcg}
P.~Meessen, D.~Mitsios and T.~Ort\'{\i}n,
JHEP \textbf{12} (2022), 155
[\arxiv{2203.13588} [hep-th]].





\end{thebibliography}
\end{document}